\documentclass[conference]{IEEEtran}
\IEEEoverridecommandlockouts
\usepackage{cite}
\usepackage{amsmath,amssymb,amsfonts}
\usepackage{algorithmic}
\usepackage{graphicx}
\usepackage{textcomp}
\usepackage{xcolor}
\usepackage{hyperref}
\usepackage{listings}
\usepackage{xcolor}
\usepackage{lipsum}
\usepackage{booktabs}

\usepackage[caption=false,font=footnotesize,labelfont=sf,textfont=sf]{subfig}

\makeatletter
\newcommand{\linebreakand}{%
  \end{@IEEEauthorhalign}
  \hfill\mbox{}\par
  \mbox{}\hfill\begin{@IEEEauthorhalign}
}
\makeatother

\lstdefinestyle{orionpython}{
  language=Python,
  basicstyle=\ttfamily\footnotesize,
  numbers=none,                     
  showstringspaces=false,
  columns=fullflexible,
  keepspaces=true,
  breaklines=true,
  frame=single,
  framerule=0.3pt,
  framesep=2pt,                     
  keywordstyle=\bfseries\color{blue!70!black},
  commentstyle=\itshape\color{gray!70!black},
  stringstyle=\color{teal!60!black},
  captionpos=b,                     
  belowcaptionskip=20\baselineskip,    
}

\begin{document}

\title{Network and Compiler Optimizations for \\Efficient Linear Algebra Kernels in \\Private Transformer Inference\\
}
\author{%
\IEEEauthorblockN{Karthik Garimella\textsuperscript{*}}
\IEEEauthorblockA{New York University \\ kg2383@nyu.edu}
\and
\IEEEauthorblockN{Negar Neda\textsuperscript{*}}
\IEEEauthorblockA{New York University \\ nn231@nyu.edu}
\and
\IEEEauthorblockN{Austin Ebel\textsuperscript{*}}
\IEEEauthorblockA{New York University \\ abe5240@nyu.edu}

\IEEEauthorblockN{}
\and
\IEEEauthorblockN{Nandan Kumar Jha}
\IEEEauthorblockA{New York University \\ nj2049@nyu.edu}
\and
\IEEEauthorblockN{Brandon Reagen}
\IEEEauthorblockA{New York University \\ bjr5@nyu.edu}
\and
\IEEEauthorblockN{}
\thanks{\textsuperscript{*}Equal contribution.}
}

\maketitle

\begin{abstract}
Large language model (LLM) based services are primarily structured as client-server interactions, with clients sending queries directly to cloud providers that host LLMs. This approach currently compromises data privacy as all queries must be processed in the cloud and in the clear. Fully Homomorphic Encryption (FHE) is a solution to this data privacy issue by enabling computations directly upon encrypted queries. However, running encrypted transformer inference is challenging as programmers must map standard kernels to the constrained instruction set provided by FHE. In this work, we explore implementations of linear algebra kernels needed for transformer inference in FHE and understand how network optimization can help mitigate FHE costs while remaining performant.

We leverage the Orion PyTorch to FHE framework to benchmark several linear algebra kernels in order to profile two linear transformation methods, packed row and BSGS, and find that BSGS outperforms packed row methods by up to $13.7 \times$ at transformer-level scales. We also incorporate network-level pruning strategies that reduce FHE runtimes of feed forward layers by up to $11.46\times$. Furthermore, we extend Orion to include ciphertext-ciphertext matrix-matrix products, a key component in the self-attention blocks. Finally, we perform a roofline analysis of FHE primitives and encrypted linear transformations and find that (SIMD encoded) implementations are memory-bound with primitives having roughly $0.1$ integer operations per byte of DRAM traffic. These findings illustrate the need for exploring alternative encoding schemes and models of computation within CKKS to unlock scalable private transformer inference. We conduct all experiments using the Orion framework which can be found at: \href{https://github.com/baahl-nyu/orion}{https://github.com/baahl-nyu/orion}.
\end{abstract}


\section{Introduction}
Today, an increasing number of applications and services are offered through cloud-based computing, which rely heavily on user data in order to be performant~\cite{meta-genai-infra-2024}. For AI-based services, large language models (LLMs) have become the de facto interface and help users understand code, learn new topics, and even provide health diagnostics~\cite{google-med-gemini-2024}.  However, the current computing paradigm requires users to upload their sensitive data directly to the cloud, and this server-side processing potentially exposes user data to unchecked processing, collection, or even data breaches. Fully Homomorphic Encryption (FHE) is a solution to this problem that allows for computation to be performed directly on encrypted data without the need for decryption~\cite{fhe}. Using FHE, it is possible to still benefit from state-of-the-art models without compromising on data privacy. 

While a promising solution to data privacy, FHE faces two major problems. First, FHE is slow given that the underlying  \textit{implementation} necessitates modular arithmetic over large integer polynomials with coefficients on the order of thousands of bits~\cite{ckks}. Second, FHE is hard to program since the programming \textit{interface} only allows for SIMD addition, SIMD multiplication, and cyclic rotation of encrypted floating point vectors~\cite{rns-ckks}. To address compute, several ASIC accelerators have been developed to bridge the gap between ciphertext and plaintext latencies~\cite{rpu,reagen2020cheetahoptimizingacceleratinghomomorphic, craterlake, bts, ark, osiris, 10.1145/3579371.3589053}. And to address programmability, compilers have been built to lower the barrier to entry for writing FHE programs~\cite{chet, porcupine, fhelipe}. In this work, we use Orion, a state-of-the-art framework that runs PyTorch code directly in FHE, to explore both compiler and network optimizations that accelerate the core linear algebra kernels needed for encrypted language model inference~\cite{orion}.

In detail, we provide an overview of the key linear algebra kernels required for private, outsourced transformer inference. We focus on the linear transformations used in the feed forward layers and the matrix-matrix multiplications required in the self-attention block~\cite{vaswani2023attentionneed}. We describe and implement two competing methods for computing dense, linear transformations under FHE and also implement a general ciphertext-ciphertext matrix-matrix multiplication algorithm. Orion allows us to rapidly iterate, profile, and implement our linear algebra kernels using the canonical SIMD encoding.

Even with state-of-the-art implementations, we find latencies to be on the order of seconds to minutes for performing transformer-sized layers under FHE. To better understand this, we build off prior characterization papers and perform a roofline analysis of SIMD-based linear algebra kernels and find that all kernels are memory bound~\cite{roofline, decastro2021doesfullyhomomorphicencryption, over100xgpu}. Our analysis informs us that while FHE is promising, new encoding schemes must be used to mitigate the memory bottleneck as well as the latencies of encrypted linear algebra kernels. 

Concretely, our contributions are as follows:
\begin{itemize}
    \item We integrate two plaintext-ciphertext matrix-vector products, row and BSGS, and a ciphertext-ciphertext matrix-matrix kernels directly in Orion. We find BSGS to outperform row-based matrix-vector products for transformer-scale matrices by up to $13.7\times$. 
    \item We evaluate the effect of layer-level optimizations by pruning non-linear activation functions in transformer feed forward blocks, eliminating FHE bootstrapping and reducing feed forward latency by $11.46 \times$.
    \item We conduct a roofline analysis which informs us that highly-optimized FHE primitive operations are memory-bound with an arithmetic intensity of 0.1 integer operations per byte of DRAM traffic.
\end{itemize}

\section{Background}
In this section, we describe the CKKS scheme~\cite{ckks, rns-ckks}, the transformer architecture~\cite{vaswani2023attentionneed}, and the Orion framework~\cite{orion}.
\subsection{CKKS}

At a high level, the CKKS FHE scheme encrypts vectors of real or complex numbers that are a fixed power-of-two length $n$ and enables SIMD addition, SIMD multiplication, and rotation of the underlying vectors~\cite{ckks}. These properties make CKKS a suitable choice for encrypted deep learning applications, which operate over the reals. We refer the reader to Table \ref{tab:ckks_params} for relevant CKKS parameters and our notation.

\subsubsection{Packing}
In order to operate under the constraints of CKKS, we must first \textit{pack} our deep learning tensor data (i.e., kernel weights, matrices) into the slots of $n$-length vectors. We call these vectors cleartexts as they represent the original data in a flattened form. As we will show in later sections, a particular packing technique will influence not only the runtime but the overall system performance.

\subsubsection{Slot Encoding}
The process of encoding converts a cleartext vector into a plaintext polynomial that is an element of the ring $R_Q = \mathbb{Z}_Q[X]/(X^N + 1)$ where $N = 2n$ is the degree of the polynomial and $Q$ is the modulus. Each plaintext is thus a polynomial whose coefficients are in $[0, Q-1]$ with degree at most $N-1$. To be 128-bit secure and enable realistic computations, the value of $N$ is typically $2^{16}$ and $Q$ consists of roughly 1500 bits~\cite{dhbsgs}. 

We leverage the canonical slot encoding which 1) performs a variant of the inverse DFT on the elements of the cleartext, 2) multiplies each output by a large precision factor $\Delta$, and 3) finally rounds each scaled output to the nearest integer. While several encoding methods exist~\cite{Kun2023EncodingFHE}, the canonical slot encoding unlocks the SIMD properties of the underlying vectors by making use of the iDFT. Decoding reverses these steps: $ \mathsf{decode}(\mathsf{encode}(m)) \approx m$. 

A plaintext polynomial can be encrypted into a ciphertext which is a pair of polynomials in the ring $R_Q$. Encryption itself injects a small amount of noise into the ciphertext and so decrypting a ciphertext yields an approximation of the original plaintext: $\mathsf{decrypt}(\mathsf{encrypt}([m])) = [m] + [e]$, where $[e]$ is a small noise polynomial.

\subsubsection{Operations}
CKKS enables SIMD addition, multiplication, and cyclic rotation of encrypted vectors. For both addition and multiplication, one of the operands can be a plaintext rather than a ciphertext. For both ciphertext-ciphertext multiplication and rotations, an expensive key-switching process is required to ensure the resulting ciphertext can be decrypted correctly using the secret key. Key-switching is both compute and memory intensive and requires a large public key-switching key~\cite{neda2024ciflowdataflowanalysisoptimization}. Moreover, each unique rotation amount requires its own key-switching key. 

\begin{table}
\renewcommand{\arraystretch}{1.3}
\centering
\caption{Relevant CKKS parameters.}
\label{tab:ckks_params}
\begin{tabular}{cl}
\hline\hline
\textbf{Param.} & \textbf{Description} \\ \hline
$m$ & A cleartext vector of real or complex numbers with $n$ slots. \\
$[m]$ & A plaintext polynomial encoding $m$. \\
$[[m]]$ & A pair of polynomials encrypting the plaintext $[m]$. \\
$N$ & Power-of-two polynomial ring degree. \\
$n$ & Length of the vector message, typically $n = N/2$ slots. \\
$L$ & Maximum multiplicative level of a ciphertext. \\
$\ell$ & Current multiplicative level. \\
$Q$ & Initial polynomial modulus. \\
$q_i$ & Small moduli in RNS decomposition of $Q = \prod_{i=0}^{L} q_i$. \\
$\Delta$ & Scaling or precision factor. \\ \hline\hline
\end{tabular}
\end{table}

\subsubsection{RNS}
It is common to decompose the large coefficient modulus $Q$ into several machine-word sized moduli using the Residual Number System (RNS): $Q = Q_L =  \prod_{i=0}^{L} q_i$, where $L$ is known as the maximum multiplicative level~\cite{rns-ckks, rns}. RNS allows a single large polynomial to be decomposed into a series of $L+1$ limb polynomials: $R_Q \rightarrow (R_{q_0}, \ldots, R_{q_L})$ which can be operated in parallel using standard machine-word arithmetic. 

\subsubsection{Bootstrapping}
CKKS multiplications (either ciphertext-plaintext or ciphertext-ciphertext) increase the noise and the precision factor of the underlying values ($\Delta \rightarrow \Delta^2$). In order to mitigate both the noise and increased scaling factor, rescaling divides the ciphertext by one of the residual modulus $q_i$. Doing so decreases the ciphertext modulus from $Q_{\ell} = \prod_{i=0}^{\ell}$ to  $Q_{\ell-1} = \prod_{i=0}^{\ell-1}$, effectively dropping the last limb of the CKKS ciphertext. When no levels remain (i.e., our ciphertext has a modulus of $Q_0 = q_0$), we can replenish the levels of the ciphertext using an expensive bootstrapping procedure.

\subsection{Transformer Architecture}
In this section, we provide an overview of the general transformer architecture based on GPT-2 (12-layer, 125 million parameters)~\cite{vaswani2023attentionneed, radford2019language} and we use the phrase ``fully homomorphic encryption" as the input to our transformer. We note that there have been several improvements to the transformer: RoPE for token embedding~\cite{rope}, group query or multi-latent for self attention~\cite{gqa, deepseek}, gated linear activation in the FFNs~\cite{glu}, and layer norm before each block rather than after. However, in line with recent practices in FHE-enabled private LLM inference systems ~\cite{rho2025encryptionfriendly,castro2025encryptedllm}, we adopt the canonical transformer implementation in this work. 


\subsubsection{Token Embedding}
A sequence of characters must first be converted into a sequence of tokens. Our example phrase ``fully homomorphic encryption" becomes tokenized into the sequence $[2759, 3488, 46374, 15835]$ using the GPT-2 tokenizer~\cite{openai2022tiktoken}, effectively mapping chunks of words to integers. Each transformer has an associated vocabulary size (number of unique tokens), which is 50257 for GPT-2. Each integer token is then transformed into an embedding vector via a lookup table which has size $50257 \times 768$ for GPT-2 where the embedding length is known as the model's hidden dimension. Effectively, our query has been converted into a $4 \times 768$  tensor, which can now be processed by the rest of the transformer. 

We note that most private transformer inference implementations assume token embedding happens a priori~\cite{nexus, ciphergpt}. In reality, the embedding tables are parameters of the model and so embedding lookup should also be performed homomorphically~\cite{helrm}. In other words, rather than sending the encrypted embedding vectors to be processed in the cloud, the integer token values should be encrypted and sent.

\subsubsection{Self-Attention}
The self-attention layer allows the embedded tokens to interact with each other.  First, tokens are projected into a lower dimension using three projection matrices $W_q, W_k, W_v$. For GPT-2, these matrices have size $768 \times 64$ and the projections themselves produce three tensors $Q, K, V$ of size $4 \times 64$. The $4 \times 4$ attention score is calculated by a matrix multiplication $Q K^T$ and each row is normalized using softmax. The projected tokens $V$ undergo a linear combination with the normalized attention scores to produce a tensor of size $4 \times 64$. And finally, a layer normalization is applied over the projected dimension. For GPT-2, there are 12 self-attention blocks that are computed in parallel (known as the 12 heads) and each $4 \times 64$ output is then concatenated to produce a tensor of size $4 \times 768$. 

Note that two types of CKKS matrix-matrix products must occur within self-attention. Since each projection matrix is a trained parameter of the model owner, performing these projections is a ciphertext-plaintext matrix-matrix multiplication. On the other hand, the attention score calculation ($QK^T$) is a ciphertext-ciphertext matrix-matrix multiplication. In this paper, we study the latter and treat ciphertext-plaintext matrix-matrix multiplication as a series of linear transformations (similar to those in the feed-forward block). 

\subsubsection{Feed Forward Network (FFN)}
The feed-forward layer simply processes the embedding tokens through an MLP: an up-projection $W_{1}$, an element-wise activation function, and a down projection $W_2$. GPT-2 uses the GeLU activation function and the projection factor is $4$ so $W_1 \in \mathbb{R}^{768,3072}$ and $W_2 \in \mathbb{R}^{3072,768}$. With respect to each token, the FFN performs two plaintext-ciphertext matrix-vector products where the matrix is unencrypted. As we will see, several packing strategies exist for performing these linear transformations with varying homomorphic and system costs. Similar to the self-attention blocks, each FFN also includes a layer normalization.

\subsection{Orion}



    
We leverage, Orion, a framework for running deep neural networks written in PyTorch under FHE with minimal code changes. Orion abstracts away many of the low-level FHE details while maintaining state-of-the-art performance~\cite{orion}. For example, Listing \ref{lst:orion-ffn} shows an implementation of the FFN from a transformer block within Orion. 
\begin{lstlisting}[style=orionpython,
  caption={Feedforward network (FFN) implementation in Orion.},
  label={lst:orion-ffn},
  linewidth=\columnwidth]
import orion.nn as on

class FFN(on.Module):
    def __init__(self, dh):
        self.w1 = on.Linear(dh, dh*4)
        self.gelu = on.GeLU(degree=127)
        self.w2 = on.Linear(dh*4, dh)

    def forward(self, x):
        x = self.w1(x)
        x = self.gelu(x)
        x = self.w2(x)
        return x
\end{lstlisting}

\begin{figure*}[t]
    \centering
    \includegraphics[width=0.98\textwidth]{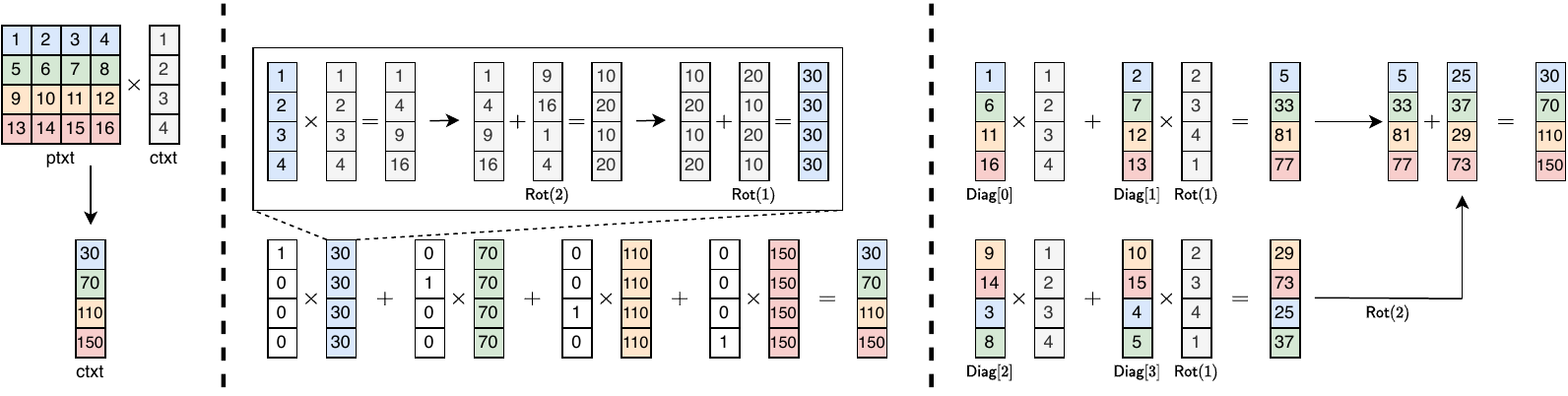}
    \caption{\textbf{Left}: The plaintext-ciphertext matrix-vector product with $n=4$ slots. \textbf{Middle}: Row methods compute each dot product \textit{within} intermediate ciphertexts, consumes two levels, and requires $O(n \log n)$ and $O(\log n)$ homomorphic operations and memory, respectively. \textbf{Right:} The diagonal-based baby-step giant-step algorithm computes dot products \textit{across} intermediate ciphertexts, consumes one level, and requires $O(\sqrt{n})$ homomorphic operations and memory.}
    \label{fig:mv}
\end{figure*}

\subsubsection{Double Hoisted Baby-Step Giant-Step}
For performant linear transformations, Orion uses a variant of the Baby-Step Giant-Step (BSGS) matrix-vector algorithm (discussed in the next section) called Double-Hoisted BSGS, which reduces the runtime of each homomorphic rotation within a matrix-vector product (see Algorithm 6 of ~\cite{dhbsgs}). While this matrix-vector algorithm is traditionally used during the bootstrapping process, Orion by default performs all linear transformations (e.g., fully-connected, convolution, permutations) using Double-Hoisted BSGS. Outside of linear transformations, Orion also adopts error-free polynomial evaluation techniques and a state-of-the-art bootstrapping protocol~\cite{dhbsgs}.

\subsubsection{Automatic Level and Scale Management}
Orion also automatically handles bootstrap placement by analyzing the network topology at the layer level and inserting bootstrap operations in order to maintain correctness while minimizing the overall network latency. Furthermore, Orion maintains the invariant that the input and output ciphertext to each layer have a precision value of exactly $\Delta$, and this property is maintained by encoding plaintexts at the appropriate scaling factor before any homomorphic operations.

\section{FFN: Matrix-Vector Products}
\label{sec:ffn}
In this section, we detail and benchmark two implementations of matrix-vector products in CKKS. In particular, these matrix-vector products are found in the feed forward blocks of transformers. We assume that the matrix (server-side) is cleartext and therefore elements of the matrix must be packed into the slots of a CKKS plaintext. This gives the server a design decision: is there an optimal packing strategy for general linear transformations?

We refer the reader to Figure \ref{fig:mv} in this section which shows a $4 \times 4$ matrix-vector product where the number of slots is also $n=4$. This is the largest possible matrix-vector product that does not overflow the slots and represents the worst-case scenario for the single ciphertext setting.

\begin{figure}[t]
    \centering
    \includegraphics[width=0.95\columnwidth]{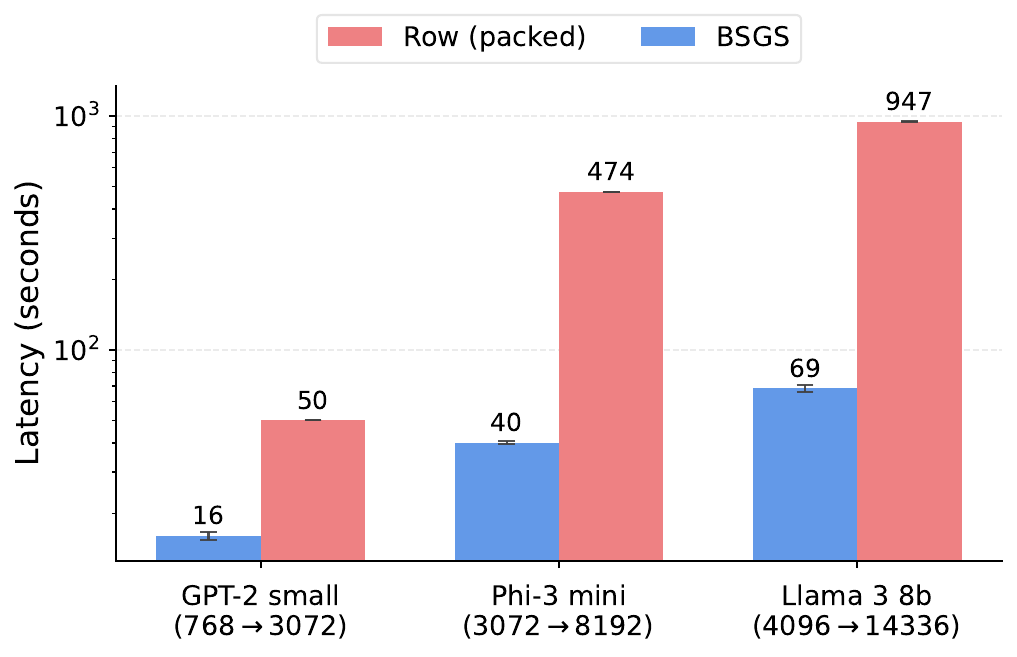}
    \caption{FHE latencies of the up projection matrix in the FFN blocks of Transformers using the (packed) row-based and BSGS algorithms. In all cases, BSGS outperforms the row method by requiring less homomorphic rotations.}
    \label{fig:matvec_comparison}
\end{figure}

\subsection{Row/Column-Based Methods}
In the row-packing method, the server packs each row of the cleartext matrix into a separate plaintext which is then multiplied by the ciphertext. This is depicted in the middle of Figure \ref{fig:mv} for the first row of the matrix. After the multiplication, the partial products across the four slots are summed up using a rotation-and-summation algorithm that has a logarithmic number of rotations with respect to $n$. 

In this case, two sequential rotations and additions are needed to sum up the partial products into each slot. This homomorphic dot product must be repeated for each row in the matrix. After each dot product has been computed, the resulting values must be consolidated into a single ciphertext via another series of multiplications (with a binary mask). 

Overall, this process requires $O(n \log n)$ rotations ($\log n$ for each of the $n$ rows) and consumes two multiplicative levels in order to compute the matrix-vector product. On the other hand, the only rotations required are powers of two meaning that this method requires generating and sending only the power-of-two rotation keys to the server.

An optimization to the row method is to pack \textit{multiple} rows of the cleartext matrix into a single plaintext if more than one row can fit into the slots. Dot products can then be computed in parallel (we refer to reader to Gazelle for more details~\cite{gazelle}). We implement this packed variant in Orion to compare against the default, baby-step giant-step method, which we now discuss.

\subsection{Diagonal Methods}
Diagonal-based matrix-vector products compute dot products across intermediate ciphertexts rather than within intermediate ciphertexts. In this case, the server packs the generalized (wrapping) diagonals of the cleartext into the slots of CKKS plaintexts and multiplies each plaintext with a ciphertext that has been aligned using a rotation. The most straightforward diagonal method (known as Halevi-Shoup~\cite{diag}) computes the matrix-vector product as $\sum_{i=0}^{n-1} \mathsf{diag}[i] \times \mathsf{Rot}(x, i)$. For example, $\mathsf{diag}[1] = [2,7,12,13]$ and $\mathsf{Rot}(x, 1) = [2,3,4,1]$ in Figure \ref{fig:mv} (right). This method requires $O(n)$ distinct rotations and rotation keys, but enables cheaper, hoisted rotations~\cite{hoist} and only consumes one multiplicative level. Given that bootstrapping ultimately consumes most cycles, we find diagonal methods to be beneficial for minimizing multiplicative level consumption and bootstrap operations. 

The baby-step giant-step algorithm (BSGS) optimizes the diagonal method by decomposing the matrix-vector product into a series of baby step rotations ($n_1$) and giant step rotations ($n_2$) such that $n_1 n_2 = n$ and $n_1 = n_2 = O(\sqrt{n})$. BSGS is visualized in Figure \ref{fig:mv} in the $n=4$ example where $n_1 = n_2 = 2$. This optimization reduces the distinct rotations and rotation keys to $O(\sqrt{n})$ by pre-rotating (for free) cleartext diagonals before packing. While BSGS is commonly used for the linear transformations within bootstrapping, Orion by default also uses BSGS (in particular, Double-Hoisted BSGS) to compute all matrix-vector products. 

\subsection{Comparison of Row and BSGS MV Products}

Figure \ref{fig:matvec_comparison} shows the difference in runtime between the (packed) row method and the Double-Hoisted Baby-Step Giant-Step algorithm for performing the first projection in the FFN block of Transformers. We average results over three runs and choose three Transformer blocks of increasing FFN sizes: GPT-2 (small), Phi-3 (mini), and Llama 3 (8B)~\cite{radford2019language, phi3mini, llama3} with their matrix sizes listed in Figure \ref{fig:matvec_comparison}. 

In all cases, the Double-Hoisted BSGS algorithm outperforms the (packed) row method and as the problem size increases, so does the speedup from using BSGS. This is because the row method scales poorly as the matrix sizes get larger: less rows can be packed into a single plaintext and there are more intermediate ciphertexts which require the rotation-and-summation algorithm. As an example, the Llama 3 FFN requires a total of 24579 rotations for the row method, while only requiring 272 rotations for BSGS.

\subsection{Impact of FFN Activation Pruning}

In this section, we examine the effect of network-level optimizations on FHE runtimes using GPT-2 (small). Following ~\cite{jha2024aero}, we first remove the element-wise GeLU nonlinearity from all the FFNs. And after pruning GeLU, we then merge the two consecutive linear layers into a single linear layer resulting in an FFN that is purely linear. In particular, the up-projection \(W_1 \in \mathbb{R}^{768, 3072}\) and down-projection \(W_2 \in \mathbb{R}^{3072, 768}\) are merged into a single linear layer, \(W_{\text{ffn}} \in \mathbb{R}^{768, 768}\), which results in an 8$\times$ reduction in FFN FLOPs. 

To examine the impact of FFN activation pruning, and further merging of the consecutive linear layers into a single linear layer, we trained the GPT-2  125M models from scratch on 2.1B training tokens from CodeParrot ~\cite{codeParrot} dataset with a 128 context window size. As shown in Table ~\ref{tab:ffn_performance}, while there is a noticeable increase in perplexity when GeLU activations are removed from all the FFN layers, the merged FFN does not result in any additional loss in perplexity.

The increases in perplexity can be mitigated by techniques such as MSE-loss based knowledge distillation ~\cite{li2023mpcformer} or entropy regularization ~\cite{jha2024aero,jha2025entropy}.  Moreover, in a different architectural setting, FFNs without element-wise nonlinearity (gated linear units) have shown  promising results, compared to the FFN with conventional nonlinear activations in LLMs ~\cite{pearce2025bilinear,glu}.

\begin{table}[htbp]
\centering
\caption{Performance comparison for FFN activation pruning.}
\label{tab:ffn_performance}
\begin{tabular}{@{}lccc@{}}
\toprule
FFN configurations & Perplexity & \#FFN FLOPs & Runtime (s) \\
\midrule
FFNs w/ GeLU & 2.688 & 14.5B & 62.21 \\
FFN w/o GeLU & 3.376 & 14.5B & 14.86 \\
Merged FFN w/o GeLU & 3.342 & 1.8B & 5.43 \\
\bottomrule
\end{tabular}
\end{table}

Furthermore, we run each configuration (averaged over 3 runs) under Orion and report the (per-layer) latency of the FFN block in Table ~\ref{tab:ffn_performance}. First, we remove GeLU (approximated as a 127-degree polynomial) which reduces the overall multiplicative level and removes the bootstrap required before GeLU. Moreover, we merge the two projection matrices reducing the latency from the original 62.21 seconds to 5.43 seconds.  

\begin{figure}[t]
    \centering
    \includegraphics[width=0.95\columnwidth]{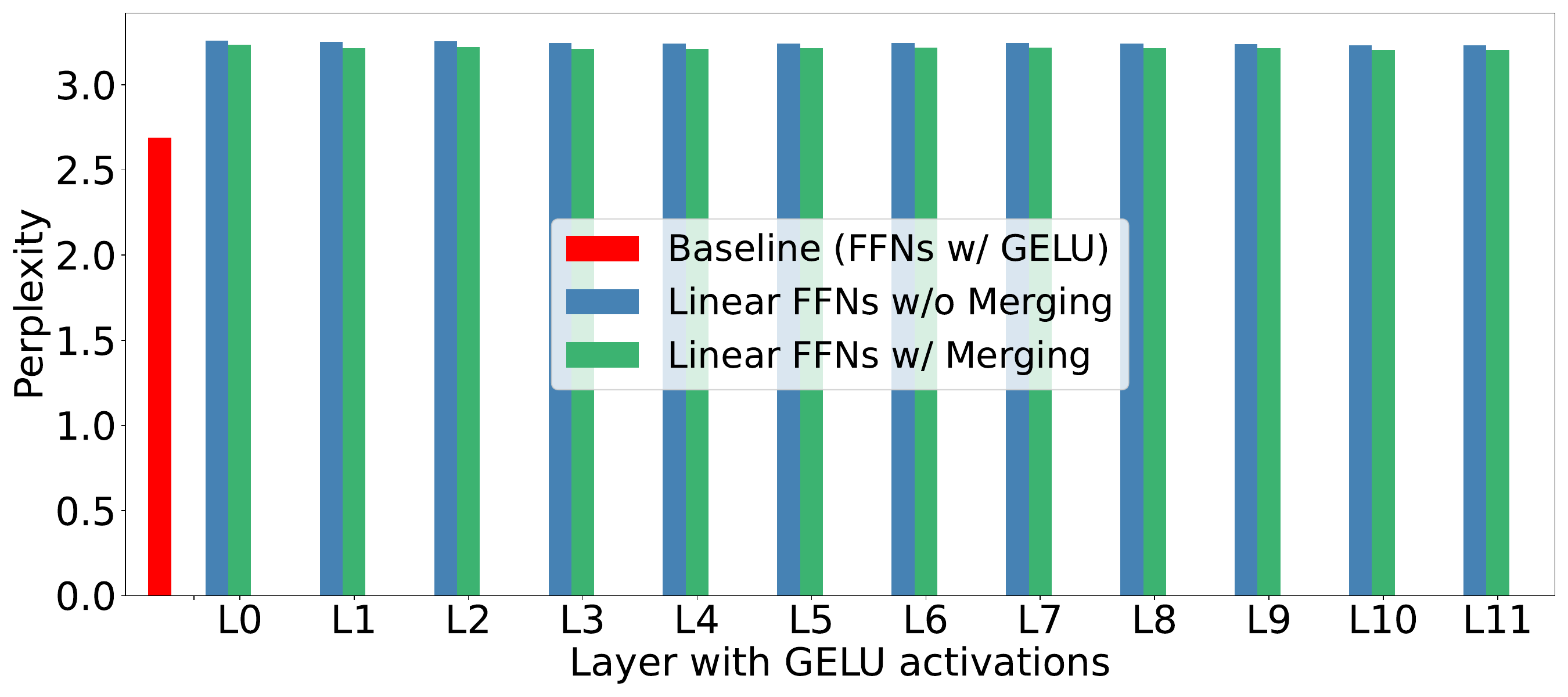}
    \caption{Perplexity comparison when GELU is removed from all but one FFN, and with their impact when linear FFNs are merged into a single linear layer. Merging linear FFNs does not incur any additional performance loss.}
    \label{fig:ffn_activation_one_layer}
\end{figure}

Finally, we perform a more fine-grained analysis to investigate the impact of FFN activation pruning on a per-layer basis. Specifically, we remove the GeLUs from all but one of the FFNs and perform merging on all other FFNs into a single linear layer.  As shown in Figure \ref{fig:ffn_activation_one_layer}, we observe that  merging linear FFNs does not incur any performance loss, which remains consistent across each merging configuration.

Note that reducing the overheads of FFNs by pruning the intervening activation, and further merging the up- and down-projections into a single layer, is a crucial step for reducing the end-to-end runtime, particularly when LLMs are operating in a smaller context window regime. In particular for the GPT-2 125M model, FFN FLOPs dominate over self-attention FLOPs for context window size smaller than 2048 ~\cite{jha2024aero}. For instance, at a 256 context window size, FFN FLOPs constitute 64\% of total network FLOPs~\cite{jha2024aero}.

\section{Self-Attention: Matrix-Matrix Multiplication}
\label{sec:sa}

\begin{figure*}[t]
    \centering
    \includegraphics[width=0.98\textwidth]{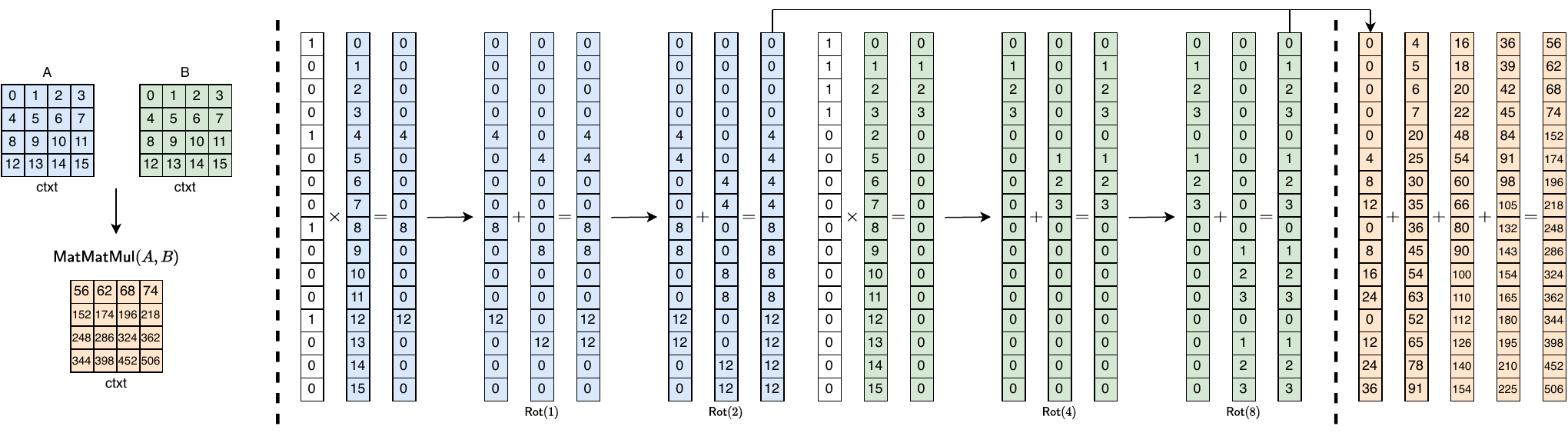}
    \caption{Ciphertext-Ciphertext Matrix-Matrix multiplication for $\sqrt{n} \times \sqrt{n}$ matrices when $n=16$ slots. Matmul is computed as a sum of outer products between each column and row of $A$ and $B$, respectively. Each column (row) is extracted via a binary mask from the ciphertexts and replicated using the rotation-and-summation algorithm. This method requires two multiplicative levels and $O(\sqrt{n})$ rotations to correctly align the outer products.}
    \label{fig:mm}
\end{figure*}

In this section, we describe the implementation of matrix-matrix multiplication in CKKS, as used in the attention layers of LLMs. We adopt the row-packing method introduced in \cite{secmatmul, aikata}.
Consider two matrices $A \in \mathbb{R}^{n \times m}$ and $B \in \mathbb{R}^{m \times p}$. The matrix product \(AB\) can be expressed as a summation of outer products between the columns of \(A\) and the corresponding rows of \(B\) as $\sum_{i=0}^{p-1} \mathsf{A_i} \otimes \mathsf{B_i}$, where $A_i$ denotes the \textit{i-th} column of \(A\) and $B_i$ denotes the \textit{i-th} row of \(B\).

Since CKKS supports only element-wise vector operations and rotation, we must express the outer product using these primitives. We assume square matrices of size $d \times d$, where $d$ is a power of two and $d \le 128$, allowing the entire matrix to be packed into a single CKKS ciphertext with $n=2^{15}$ slots. A simple row-major encoding maps the $d \times d$ matrix to a vector of dimension $2^{15}$, with the $d^2$ matrix elements placed in the first $d^2$ slots.
We denote the encoded and encrypted row-major polynomials as $A_f$ and $B_f$.
To compute the matrix-matrix multiplication, $d$ outer products must be computed. In each iteration, one column of $A$ and one row of $B$ must be extracted from their respective ciphertexts. As illustrated in Figure~\ref{fig:mm}, this extraction is achieved by multiplying each encoded matrix with a binary mask, extracting the desired column or row, therefore; each extraction requires a plaintext-ciphertext multiplication. We denote the resulting extracted polynomials as $A_{col}$ and $B_{row}$. 
After extraction, each vector must be replicated across the entire polynomial to prepare for element-wise  multiplication. This step is done using $\log 2d$ rotation and additions, as show below:

\[
\begin{aligned}
\mathbf{A}_{\text{col}} &= \mathbf{A}_{\text{col}} + \sum_{i=0}^{\log_2 d} \mathbf{A}_{\text{col}}.\text{roll}(2^i) \\
\mathbf{B}_{\text{row}} &= \mathbf{B}_{\text{row}} + \sum_{i=0}^{\log_2 d} \mathbf{B}_{\text{row}}.\text{roll}(d\times 2^{i})
\end{aligned}
\]

\begin{figure}[t]
    \centering
      \includegraphics[width=0.95\columnwidth]{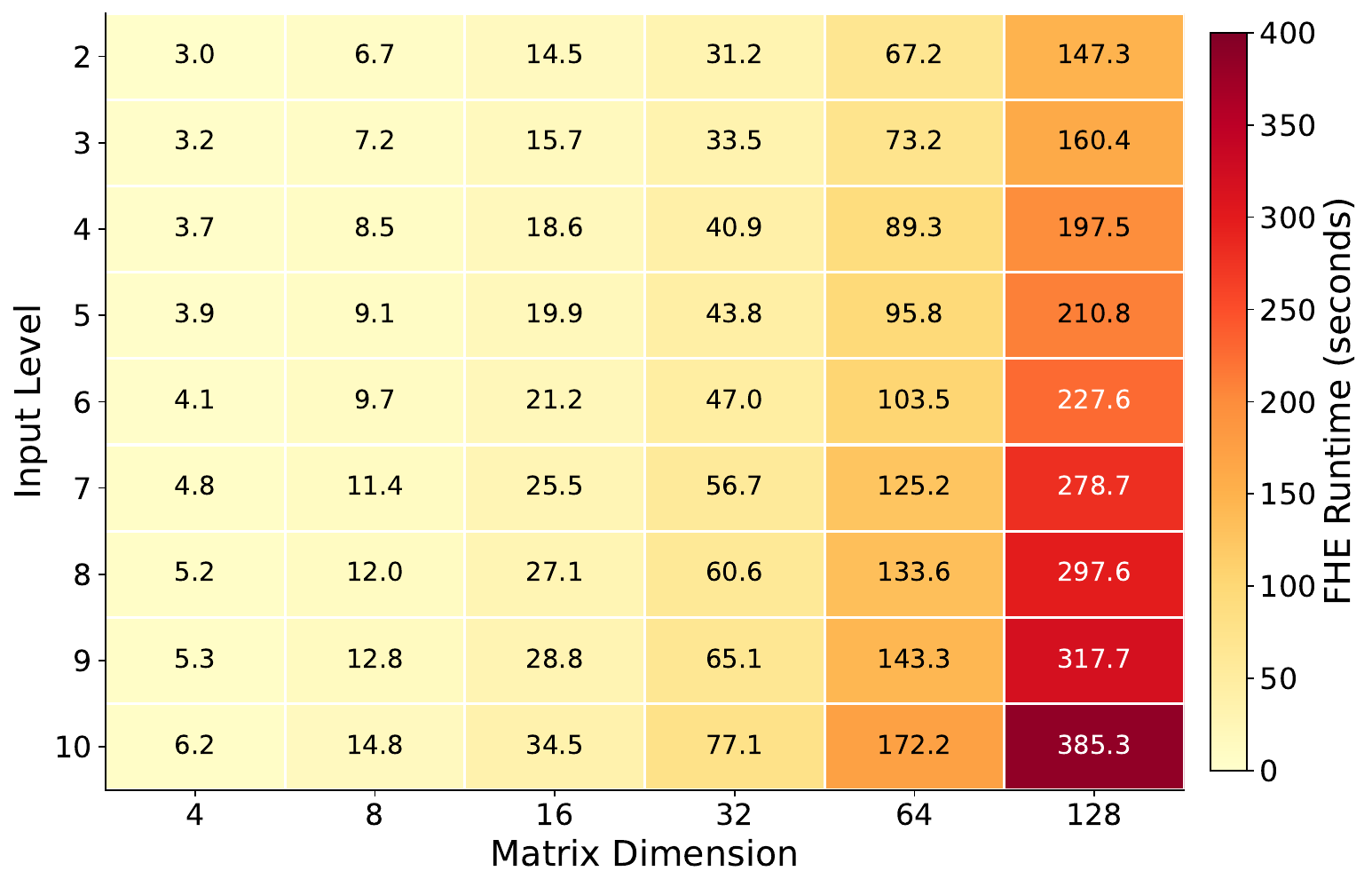}
    \caption{Ciphertext-Ciphertext Matrix-Matrix multiplication latencies as a function of input level and matrix dimension. Each matrix multiply operates on power-of-two square matrices. }
    \label{fig:mm_plot}
\end{figure}

The result of the element-wise multiplication between $A_{col}$ and $B_{row}$ is a dot product, which is stored in a row major format in the output polynomial.
In addition to the replication rotations, additional rotations are required for each outer product to align the valid coefficients at the top of the polynomials. Specifically, $B_{row}$, is rotated by $d\times j$ and $A_{col}$ by $j$, where $0<jd$, depending on the current iteration. These rotations ensure that the extracted values are correctly positioned for accumulation into the final result. 
Each step of computing the partial products in this algorithm requires $\mathcal{O}(\log_2d)$ rotations on a single ciphertext, one PT-CT multiplication, and one CT-CT multiplication. Overall, the entire computation requires $2 \times d$ PT-CT multiplications, $d$ CT-CT multiplications and $2\times d(1+log_2d)-2$ rotations.
Figure~\ref{fig:mm_plot} shows the runtime of matrix-matrix multiplication using the mentioned algorithm and the Orion~\cite{orion} framework. The x-axis represents the matrix dimension, and the y-axis indicates the input level of the ciphertext that defines the maximum multiplicative level of the ciphertext, which determines its maximum multiplicative length. This algorithm requires a multiplicative level of 2; therefore the input level must be at least 2 to produce a correct result. As shown, increasing either the matrix dimension or the input level leads to higher runtimes. For GPT-2 with a sequence length of 64, the matrix-matrix computation takes between $67.2$ and $172.2$ seconds depending upon the level at which the computation is performed.

\section{Roofline Analysis of FHE}
Looking back at both Sections \ref{sec:ffn} and \ref{sec:sa}, we see that SIMD-based linear algebra kernels for encrypted transformer models still operate on the order of seconds to minutes. For example, FFNs take 16 seconds per token and on average 2 minutes per attention matrix calculation for GPT-2 (small). And for ``small" language models such as Llama 3 with 8 billion parameters, these kernels take roughly 15 and 5 minutes, respectively. We note that these latencies are for a \textit{single block} for each model. To this end, we now analyze the arithmetic intensity of core FHE primitives to determine their performance bottlenecks. 


\begin{figure*}[t]
    \centering
    \includegraphics[width=\textwidth]{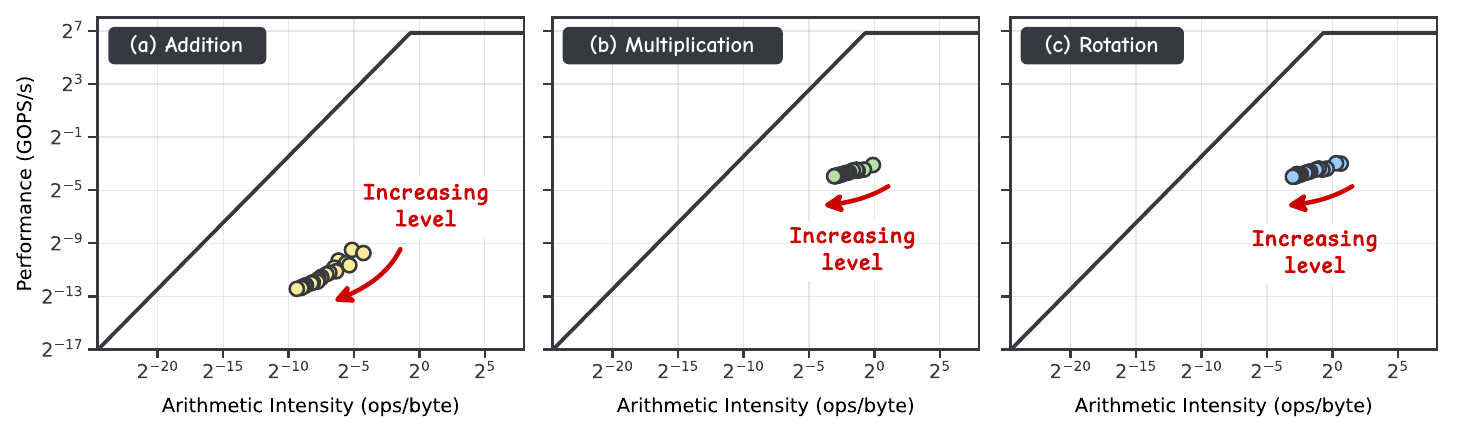}
    \vspace{-15px}
    \caption{Rooflines for FHE primitives: ciphertext-ciphertext addition, ciphertext-ciphertext multiplication (with relinearization), and ciphertext rotation. All three primitives are primarily memory-bound and have decreasing arithmetic intensities with respect to multiplicative level. Both multiplication and rotation have similar characteristics as key-switching is the dominant component for both operations. We use 64-bit integer operations per byte for arithmetic intensity.}
    \label{fig:three-side-wide}
\end{figure*}

\subsection{Overview of Rooflines}
Performance improvements in any kernel primarily come from three sources: increasing DRAM bandwidth, increasing total compute (through more cores or higher frequency), and with algorithmic optimizations that promote smarter data caching and reuse. The latter requires navigating a complex space of optimizations (e.g., loop tiling, vectorization, etc.), each with their own trade-offs and implementation strategies. The roofline model intuitively quantifies which of these approaches will yield the greatest speedup. Here, applications are characterized by their arithmetic intensity, which traditionally refers to the number of floating-point operations performed per byte of DRAM traffic. If an application has low arithmetic intensity, then very few operations are performed compared to data transferred. This imbalance often leaves CPU cores idle waiting for data to arrive, and therefore increasing compute will provide only marginal performance improvements. In contrast, high arithmetic intensity operations benefit directly from additional compute resources since memory bandwidth is no longer the limiting factor. 

Our roofline models are shown in Figure \ref{fig:three-side-wide} and  consist of two regions: a sloped line representing the memory bandwidth limit and a horizontal line representing peak computational throughput. When combined, they represent the peak attainable performance for a given CPU. Applications that fall below the sloped region are memory-bound and limited by bandwidth, while those approaching the horizontal limit are compute-bound and limited by the processor's peak performance. An ideal program sits at the elbow or ``ridge point" of the graph, achieving peak arithmetic throughput without being bottlenecked by memory transfers.

\subsection{Modifications for FHE} Directly applying a roofline analysis to FHE is challenging given that virtually no floating point computations are being performed. In this work, we modify the arithmetic intensity metric to instead measure \textit{integer} operations per byte of DRAM traffic. Here, an integer operation consists of $64$-bit \texttt{ADD}, \texttt{SUB}, \texttt{MUL}, or \texttt{DIV} instructions operating on general-purpose registers, excluding stack operations and immediate values. We use Intel PIN~\cite{pin} for dynamic binary instrumentation to count these operations at runtime, inserting markers around FHE kernels to isolate their operation counts. DRAM traffic is measured using hardware performance counters that track memory controller CAS (Column Address Strobe) events. We use OpenFHE~\cite{openfhe} as the CKKS backend and set the thread count for each experiment to $48$ in order to make full use of our machine’s available compute. For simplicity, we disable vectorization in OpenFHE and limit the maximum core frequency to 2.4 GHz. We also set the polynomial degree to be $N = 2^{16}$ giving us a slot count of $n = 2^{15}$. 

\subsection{FHE Primitives Analysis}

Figure \ref{fig:three-side-wide} visualizes roofline plots for each primitive FHE operation. Here, we sweep the multiplicative level from $1$ to $20$, measuring latency, integer operation counts, and total DRAM traffic. Following a traditional client-server setup, we assume that any input data and rotation keys are loaded directly from disk, rather than stored on-chip. Peak compute is calculated as the product of CPU frequency and active core count. Measured performance (in GOPS/s) is then calculated by dividing the number of integer operations by the observed latency. As shown in Figure \ref{fig:three-side-wide}a, ciphertext-ciphertext addition is both memory-bound and exhibits a low arithmetic intensity. This is not surprising as the only operations needed are modular additions over the (limbed) polynomials within each ciphertext. As the level increases, the arithmetic intensity decreases. A similar trend is seen for both multiplication and rotation (Figure \ref{fig:three-side-wide}b and c). In both operations, the key-switching procedure is the most time (and memory) consuming step, meaning that both kernels share similar roofline characteristics.

In line with prior work~\cite{decastro2021doesfullyhomomorphicencryption, over100xgpu}, we find that FHE primitives are primarily memory-bound, \textit{especially as the ciphertext level increases}. Despite using OpenFHE as our cryptographic backend, these trends persist in other libraries that adopt traditional key-switching techniques ~\cite{ckks,lattigo, seal}. Our observations inform us that alternative encoding schemes and models of computation for FHE are necessary to reduce overall latency~\cite{Ju_2024, bae2025fasthomomorphiclinearalgebra}.




\section{Related Works}
\subsection{FHE Compilers}
CHET and EVA were among the first FHE compilers to target machine learning workloads lowering to the SEAL library~\cite{chet, eva, seal}. These works built an intermediate representation, optimized data packing layouts, and introduced waterline rescaling. Additionally, FHE frameworks such as TenSEAL and nGraph-HE target deep learning applications but these frameworks do not include automatic bootstrap placement~\cite{tenseal, ngraphhe}.

More recently, compilers such as Dacapo, HeLayers, Fhelipe, and Orion ~\cite{dacapo, helayers, fhelipe, orion} include support for automatically inserting bootstrap operations into deep learning applications to enable deep computations such as ResNet-50 and YOLO networks~\cite{resnet, yolo}. 

\subsection{Private Transformer Inference}
There are several prior FHE-based private transformer inference papers. NEXUS was the first purely FHE-based implementation that used a ciphertext compression and expansion technique to reduce the amount of communication and outperforms hybrid protocols in terms of inference cost~\cite{nexus}. \cite{castro2025encryptedllm} open-sourced a GPU implementation of GPT-2 and developed non-linear kernels for GeLU, Layernorm, and Softmax.  ~\cite{10.1145/3643651.3659893} introduced several packing strategies for matrix-matrix multiplications and mitigated LayerNorm precomputing the mean and variance. In PowerSoftmax, stable training is explored~\cite{zimerman2024powersoftmaxsecurellminference}.

THOR~\cite{thor} reduces matrix-matrix multiplication runtime by utilizing a diagonal packing strategy, akin to the diagonal based matrix-vector products that we explored in this paper. Powerformer~\cite{powerformer} focuses mainly on the nonlinear activation functions present in Transformers exploring replacements for Softmax that are more privacy-friendly. ~\cite{rho2025encryptionfriendlyllmarchitecture} et al. reduce the runtime of ciphertext-ciphertext matrix-matrix multiplication by utilizing LoRA~\cite{hu2021loralowrankadaptationlarge}. Finally, Tricycle is a new private transformer inference framework that utilizes tricyclic encoding that immediately enables batched matrix-matrix multiplication using only one multiplicative level~\cite{tricycle}.

\section{Conclusion}
In this work we leveraged Orion, a framework that runs PyTorch models under FHE, to implement several linear algebra kernels required for private transformer inference. In particular, we compared two competing matrix-vector product algorithms (packed row and BSGS) and observed that for transformer-scale linear transformations, BSGS outperformed the packed row method by up to $13.7 \times$. Furthermore, we incorporated network level optimizations which reduced FFN runtime by $11.46 \times$ and also expanded Orion to include a general row-column based matrix-matrix multiplication algorithm, a key kernel used in the self-attention mechanism. Finally, we performed a roofline analysis of FHE operations and found all primitives (addition, multiplication, and rotation) to be memory bound at around 0.1 integer operations per byte of DRAM traffic. 

Going forward, we believe that for FHE-based transformer inference to be practical, it is necessary to research new encoding schemes and computational methods under FHE that move past the canonical slot encoding technique. Doing so would mitigate the memory pressure faced by current FHE applications and allow for scalability to larger problem sizes. We hope our analysis in this paper guides the research community to think holistically about FHE system design, from the core cryptography to the computer and network architecture design.

\section*{Acknowledgements}
\noindent
This work was supported in part by Graduate Assistance in Areas of National Need (GAANN). The research was developed with funding from the NSF CAREER award \#2340137 and DARPA, under the Data Protection in Virtual Environments (DPRIVE) program, contract HR0011-21-9-0003. Reagen and Ebel received a gift award from Google. The views, opinions, and/or findings expressed are those of the authors and do not necessarily reflect the views of sponsors.

\bibliographystyle{IEEEtran}
\bibliography{references}

\vspace{12pt}
\end{document}